\documentclass[]{emulateapj}

\def\HI{\hbox{H~$\scriptstyle\rm I\ $}}

\def\CIV{\hbox{C~$\scriptstyle\rm IV\ $}}

\def\SiIV{\hbox{Si~$\scriptstyle\rm IV\ $}}

\def\mg{{\rm MgII\,\,}}

\def\spose#1{\hbox to 0pt{#1\hss}}
\def\lta{\mathrel{\spose{\lower 3pt\hbox{$\mathchar"218$}}
     \raise 2.0pt\hbox{$\mathchar"13C$}}}
\def\gta{\mathrel{\spose{\lower 3pt\hbox{$\mathchar"218$}}
     \raise 2.0pt\hbox{$\mathchar"13E$}}}
\newcommand{\be}{\begin{equation}}
\newcommand{\ee}{\end{equation}}

\newcommand{\mincir}{\raise
-2.truept\hbox{\rlap{\hbox{$\sim$}}\raise5.truept
\hbox{$<$}\ }}
\newcommand{\magcir}{\raise
-2.truept\hbox{\rlap{\hbox{$\sim$}}\raise5.truept
\hbox{$>$}\ }}
\newcommand{\minmag}{\raise-2.truept\hbox{\rlap{\hbox{$<$}}\raise
6.truept\hbox
{$>$}\ }}

\lefthead{Porciani, Viel \& Lilly }
\righthead{Strong \mg systems in QSO and GRB spectra}

\begin{document}

\title{Strong \mg systems in quasar and gamma-ray burst 
spectra} \author{Cristiano Porciani \altaffilmark{1}, Matteo Viel
\altaffilmark{2,3}, Simon J. Lilly \altaffilmark{1}}
\altaffiltext{1}{Institute for Astronomy, ETH Z\"urich, 8093 Z\"urich,
Switzerland; porciani@phys.ethz.ch; simon.lilly@phys.ethz.ch.}
\altaffiltext{2}{Institute of Astronomy, Madingley Road, Cambridge CB3
0HA, United Kingdom} \altaffiltext{3}{INAF - Osservatorio Astronomico
di Trieste, Via G.B. Tiepolo 11, I-34131 Trieste, Italy;
viel@oats.inaf.it.}

\begin{abstract}
The incidence of strong \mg systems in gamma-ray burst
(GRB) spectra is a few times higher than in quasar (QSO) spectra. We
investigate several possible explanations for this effect, including: dust
obscuration bias, clustering of the absorbers, 
different beam sizes of the sources, multiband magnification
bias of GRBs, association of the absorbers with the GRB event or the 
circumburst environment.  
We find that: 
{\it i)} the incidence rate of \mg systems in
QSO spectra could be underestimated by a factor 1.3-2 due to dust
obscuration; 
{\it ii)} the equivalent-width
distribution of the \mg absorbers
along GRBs is consistent with that observed along QSOs thus suggesting
that the absorbers are more extended than the beam sizes of the sources;
{\it iii)} on average, GRB afterglows showing more than one \mg system
are a factor of 1.7 brighter than the others, suggesting a lensing
origin of the observed discrepancy;
{\it iv)}
gravitational lensing (in different forms, from galaxy lensing to microlensing)
can bias high the counts of \mg systems along GRBs if
the luminosity functions of the prompt gamma-ray emission and of the 
optical afterglows
have a mean faint-end slope approaching -5/3 -- -2; 
{\it v)} some of the
absorbers can be associated with the circumburst environment or
produced by supernova remnants unrelated to the GRB
event itself but lying in the same star-forming region.
With the possible exception of magnification bias, it is unlikely
that one of these effects on its own can fully account for
the observed counts. However, the combined action of some of them can
substantially reduce the statistical significance of the discrepancy.
\end{abstract}

\section{Introduction}
Magnesium is an $\alpha$-process element produced by red-giant stars
and dispersed in the interstellar medium by supernova explosions and
stellar winds. 
In the redshift interval $0.3\mincir z\, \mincir 2.2$,  
the \mg doublet (2796, 2804 \AA) produces 
absorption lines in the optical spectrum of background sources. 
The strong absorption systems (with equivalent width $W>0.3$ \AA) 
are thus believed to be good tracers of metal-enriched gas
associated with galaxies 
(e.g. Bergeron \& Boisse 1991; Steidel, Dickinson 
\& Persson 1994). Typically these systems have multiple velocity components
(Churchill and Vogt 2001). 

Recent high-resolution imaging studies of quasar fields provided
evidence that strong \mg absorption is produced in patchy
gaseous envelopes, up to impact parameters of $80\, h^{-1}$ kpc,
surrounding galaxies of different morphological types (Churchill,
Kacprzak \& Steidel 2005 and references therein). The completion of
large and homogeneous quasar (QSO) samples as the Sloan Digital Sky
Survey (SDSS; York et al. 2000) 
and the Two-degree field quasar survey (2Qz; Boyle et al. 2000;
Croom et al. 2004) allowed
accurate statistical studies.  The amplitude of the cross-correlation
function between \mg systems and luminous red galaxies suggests that
absorbers with $W>1$ \AA\ are hosted within dark-matter halos with
characteristic masses of $10^{11-12} M_\odot$ and could be associated
to galactic superwinds (Bouch\'e et al. 2006). 
Using the SDSS Early Data Release,
Nestor, Turnshek \& Rao (2005) identified over 1,300 \mg doublets with 
$W>0.3$ \AA and 
measured their equivalent width distribution
over the redshift range $0.366 \le z \le 2.269$.
Similarly,
Prochter, Prochaska \& Burles (2006a) found nearly 7,000 \mg systems
with $W>1$ \AA\ in the spectra of 50,000 QSOs from the SDSS DR4.  This
corresponds to a redshift path density $dN/dz\simeq 0.24$ at
$z=1$. These results have been extended to lower redshift (and weaker column
densities) by Nestor, Turnshek \& Rao (2006). 
This study 
suggests that the gas clouds associated with weak \mg systems
are a physically distinct population from those producing strong \mg absorbers
(see also Nestor et al. 2005).

Gamma-ray bursts (GRBs) with bright optical afterglows 
can also be used as background sources. Somewhat surprisingly,
Prochter et al. (2006b, hereafter P06) identified 14 \mg systems
with $W>1$ \AA\  along 14 GRB lines of sight (for a total redshift
path of 15.5 at a mean redshift $\bar{z}=1.1$). 
This corresponds to $dN/dz=0.90_{-0.50}^{+0.83}$ (symmetrical 99\% Poisson
\footnote{Clustering of the absorbers produces
super-Poisson fluctuations, this effect will be discussed in \S\ref{dust}.}
confidence interval), 
an incidence rate that is a few
times larger than that inferred from the QSOs of the SDSS data set.
A similar (but much less statistically significant) discrepancy,
$dN/dz=0.62_{-0.49}^{+1.13}$ (with an additional $\sim 30\%$
uncertainty due to the redshift path), has been reported by Stocke \&
Rector (1997) in the optical spectra of 21 radio-selected BL Lacertae
objects.  However, recent redshift determinations (see
e.g. Sbarufatti, Treves \& Falomo 2005) showed that 1 out of the 5
strong systems found by Stocke \& Rector (1997) is associated with the
host galaxy. This reduces the cosmological incidence rate
to $dN/dz=0.49_{-0.41}^{+1.06}$.

In this work, we critically review some possible explanations
of the discrepancy between the counts of \mg systems along GRBs
and QSOs.
%and show that, even though there is not 
%a single effect that can fully account for it, the combined action of
%a number of phenomena can substantially reduce its statistical significance.
The paper is organized as follows.
In \S \ref{dust}, we estimate the importance of
dust-obscuration bias along lines of sight towards QSOs.
The effect of different beam sizes between GRBs and QSOs and the
solution proposed by Frank et al. (2006, where QSO beams are assumed to be
larger than GRB ones by a factor of 2) are discussed in \S \ref{beamsize}.
The roles of
gravitational lensing and magnification bias of GRB afterglows 
are presented in \S \ref{lensing}.
Finally, in \S \ref{association}, we discuss the 
possibility that some of the \mg systems along GRBs are produced in
the circumburst environment.
Our results are summarized in \S \ref{summary}.

\section{Dust obscuration bias}
\label{dust}

 The presence of dusty absorbers along the line of sight
 could obscure the optical light from background QSOs and produce a selection
 bias in magnitude-limited samples (Ostriker \& Heisler 1984; Heisler \& Ostriker 1988; Fall \& Pei 1993). 
 If the obscuration bias is important, radio selected QSOs (which are
 unaffected by dust) should present a larger number of absorbers on
 average.  Recent studies did not find strong evidence for dust
 reddening and extinction (Ellison et al. 2001; Akerman et al. 2005;
 Jorgenson et al. 2006).  However, radio samples are very small and
 cannot lead to definitive conclusions. Even though the number counts
 of absorbers in radio and optically selected QSOs are in agreement,
 1$\sigma$ statistical uncertainties are consistent with 60\% of
 damped Ly-$\alpha$ systems being missed in optical magnitude-limited
 surveys (Ellison et al. 2004).

Could dust obscuration bias of the QSO samples explain the discrepancy
with the incidence of \mg systems in GRB spectra? 
A few dusty systems with color excess $E(B-V)\sim 0.1$
have indeed been found (Junkkarinen et al. 2004, Ellison et al. 2006).
On the other hand, 
statistical studies 
give contrasting evidence for
dust obscuration. 
Characteristic values of $E(B-V)=0.06-0.1$
have been reported for 
ZnII and CaII systems with large $W$ 
(Vladilo \& Peroux 2005; Wild et al. 2006). 
However, Murphy \& Liske (2004) found no evidence for dust-reddening
of QSOs by foreground damped Lyman-$\alpha$ systems.
Similarly, York et al. (2006) could measure some appreciable reddening only
in QSOs with very strong \mg absorbers ($W>1.53$ \AA).
Moreover, radio and X-ray selected SDSS quasars do not appear to be 
more reddened than optically selected ones. In both cases the typical color excess is
$E(B-V)\simeq 0.01$.

All these results are puzzling and contradictory.
In order to estimate the importance of dust obscuration in the SDSS sample
of \mg absorbers we have developed the following simplified model. 
We consider a population of QSOs at $z=2.3$ with luminosities  
distributed according to the luminosity function determined by the SDSS
(Richards et al. 2006).
These QSOs will be obscured by foreground galaxies.  
To model the galaxy distribution, we use 24 mock light cones extracted
from the largest N-body simulation of the concordance cosmology
performed so far, the Millennium run (Springel et al. 2005), 
in which galaxies are associated with
dark-matter halos via semi-analytic modelling (Kitzbichler et
al. 2006). 
The past light cone of an observer (i.e. an event with
coordinates $x,y,z,t_0$) consists of all the paths of light that reach
the observer at $t_0$. Basically, it contains all the galaxies that the
observer could detect with an ideal experiment since they are connected 
by null geodesics with the observer himself.
We shoot $10^3$ random lines of sight for each light cone and count how many galaxies we find within a given
impact parameter $b=80 h^{-1}$ kpc (Churchill et al. 2005).  We then
associate a \mg absorber to an intervening galaxy probabilistically
with a covering factor $f=0.5$ (Churchill et al. 2005).  We only
consider galaxies with (unextincted) 
absolute magnitude $M_B<M_{\rm thr}$ and choose
$M_{\rm thr}$ to match a given $dN/dz$.
As a first case, 
we also assume that \mg systems contain some dust with a bimodal
distribution: a fraction $f_{\rm d}$ of them has $E(B-V)=0.1$ (with a 
Small Magellanic Cloud (SMC) extinction curve)
while all the rest has $E(B-V)=0.01$.
Dust obscuration cannot be discussed separately from gravitational lensing
that boosts the luminosity of background sources. 
For this reason, we also compute the magnification due to all intervening
mass concentrations along a given line of sight assuming that they are 
singular isothermal spheres (the resulting magnifications have a mean of 1.03
and an r.m.s. value of 0.04 but the distribution is very positively skewed,
nearly 0.8\% of the sources are amplified by more than 20\%).
\begin{figure}
\plotone{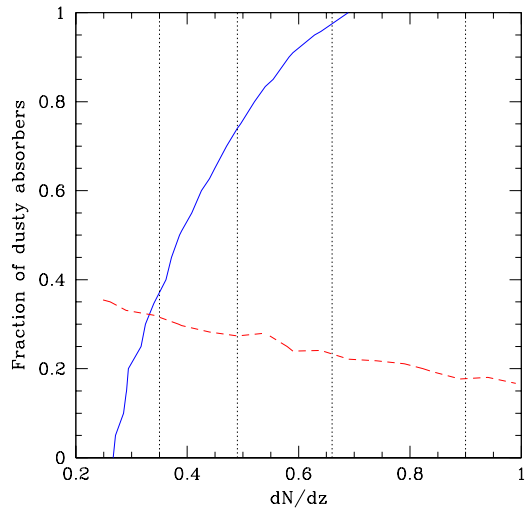}
\caption{\label{bimodal}
The underlying number density of strong \mg systems and 
the fraction of dusty absorbers that match the observed abundance of
\mg systems (solid)
and the observed fraction of absorbers along very reddened lines of
sight (dashed) in the SDSS. 
The dotted lines
mark the 0,1,2 and 3$\sigma$ Poisson fluctuations for the observed density
along GRBs.}
\end{figure}

Our results are summarized in Figure \ref{bimodal}.
The abscissa shows the true number density of the underlying \mg systems
(those revealed along GRB lines of sight), while the ordinate gives $f_{\rm
d}$: the fraction of dusty absorbers.
The solid line indicates the parameter pairs for which SDSS would observe
$dN/dz=0.24$. 
Obscuration bias can decrease the observed density of absorbers by 
a maximum factor of $\sim 3$.
In order to match the observed density of strong \mg systems in SDSS,
the number density of underlying absorbers cannot be higher than
$dN/dz\simeq 0.7$ but, in this case, all \mg system should be very dusty.
This would be in contrast with observations.
York et al. (2006) found that only $14\%$ of the \mg systems
with $W>0.3\,$ \AA\, in the SDSS catalog (and $36\%$ of those with $W>2.4$ \AA)
lie along very reddened lines of sight ($\Delta(g-i)\geq0.2$).
If this fraction amounts to $\sim 20\%$ for $W>1$ \AA,
the only way to reconcile our results with the observational data is 
the case where $dN/dz\simeq 0.33$ (corresponding to 
$M_{\rm thr}\simeq -21.65$) 
\footnote{Note that reducing either the maximum impact parameter $b$ or
the covering factor $f$ would make the absorbers' host galaxies fainter
for a given $dN/dz$.}
and $f_{\rm d}\simeq 0.33$.
In this case, SDSS would miss 16\% of the QSOs which are intrinsically brighter
than its magnitude limit (the effect of color selection is less important)  
and include 1\% of the total number 
due to magnification bias (plus dust obscuration).
Note that the median and mean halo masses of the \mg absorbers are
$3.0 \times 10^{11}$ M$_\odot$ and
$7.5 \times 10^{11}$ M$_\odot$,  
in good agreement with Bouch\'e et al. (2006).
We explicitly checked that our results do not change substantially 
by using a Milky Way 
(MW) extinction curve for the dusty absorbers.

In order to test if our results depend on the assumed distribution of
reddening, we repeated our Monte Carlo simulations assuming a
two-parameter Weibull distribution for $E(B-V)$ (for an SMC extinction
curve). 
%
%This distribution function can attain many different shapes based on 
%the value of one of its parameters. 
%
The cumulative Weibull distribution for a variable $x\geq 0$ is
$C(x)=1-{\rm exp}[-(x/\lambda)^\gamma]$ and it is fully determined by
the shape parameter $\gamma>0$ and the scale parameter
$\lambda>0$. This distribution is interesting because it can attain
many different shapes based on the value of $\gamma$.  In particular,
for $\gamma<1$ the corresponding probability density function
decreases monotonically and is convex; for $\gamma=1$ it becomes the
exponential distribution; for $\gamma>1$ it vanishes at $x=0$ and admits
a mode at $x= \lambda \,(1-1/\gamma)^{1/\gamma}$; for $\gamma=2$ it gives
the Rayleigh distribution; for $\gamma<2.6$ it is positively skewed;
for $2.6<\gamma<3.7$ it closely approximates a normal distribution;
for $\gamma>3.7$ it is negatively skewed. We find that only monotonically
decreasing distributions for $E(B-V)$ with $\gamma\sim 0.3-0.5$
are compatible with the data. 

\begin{figure}
\plotone{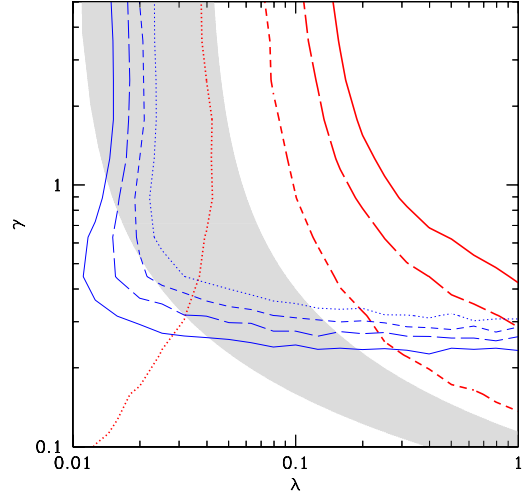}
\caption{\label{weibull}
Parameters of the reddening distribution of strong \mg systems.
Thick and thin curves respectively mark the set of parameters for which the 
observed abundance of absorbers and the fraction of very reddened lines of 
sight match 
the SDSS values. Solid, long-dashed, short-dashed and dotted lines
refer to $dN/dz=0.90, 0.66, 0.49, 0.35$ (i.e. to 0,1,2 and 3$\sigma$ Poissonian
fluctuations in the counts of GRB absorbers) respectively.
The shaded area is bounded by the loci where the median observed color
excess $\lambda\,[\ln(2)]^{1/\gamma}$
is 0.01 (lower boundary) and 0.04 (upper boundary).
Models for which thin and thick curves of the same type cross within the shaded
area are consistent with all the observational constraints.
} 
\end{figure}

In this case, there are many solutions that 
match both the density of \mg absorbers
and the fraction of absorbers along very reddened QSOs observed by SDSS
(see Figure \ref{weibull}).
However, if we also require that the median observed color excess is of
the order of 0.01 (as found in the SDSS by York et al. 2006) we have to assume
that the underlying incidence rate is $dN/dz\simeq 0.30$
(corresponding to $M_{\rm thr}\simeq -21.7$).
This increases up to $dN/dz\simeq 0.45 $ ($M_{\rm thr}\simeq -21.4$) 
if the 2Qz $E(B-V)$ value 
(0.04, Outram et al. 2001) is adopted.

In summary, we found that estimates of $dN/dz$ based on QSO spectra are likely
to be underestimated by a factor $\sim 1.3-2$ if dust-obscuration bias is important.
If $dN/dz\sim 0.35$ (0.45), we find that there is a $0.15\%$ 
(1.3\%) chance to find 14 absorbers or more 
in a redshift path of 15.5 because of random fluctuations. 
This shows that it is unlikely that dust obscuration bias fully explains
the difference in the number of \mg absorbers along GRBs and QSOs.
Note that
in our simulations, the scatter of the number of absorbers per line 
of sight is generally
larger than expected for a Poisson distribution because the host-galaxies
of the absorbers are clustered.
If, on average, there are $\bar{N}$ absorbers per line of sight,
the variance of the counts 
is $\sigma^2=\bar{N}\,(1+\bar{N}\,\bar{\xi})$ where $\bar{\xi}$ is the mean
two-point correlation function computed by averaging over pairs of points
both lying within a narrow cylinder with radius $b$ parallel to the line
of sight.
For $b=80\, h^{-1}$ kpc and a redshift interval $0.3<z<2.2$, the galaxies
hosting the absorbers have typically mean correlations of some percent 
(slightly depending on the mean galaxy luminosity). 
Whenever $\bar{N}\magcir 1$, superPoisson fluctuations are then appreciable.
A number, $X$, of independent lines of sight have to be combined together to 
obtain a redshift path of 15.5. 
This increases both the mean counts and the scatter
by a factor of  $X$ so that the total excess of the scatter with respect
to Poisson remains $1+\bar{N}\,\bar{\xi}$ as for single lines of sight.
In consequence, the scatter in the number of \mg systems
is nearly Poissonian along QSOs 
and superPoissonian along GRBs. Future data with increased statistics,
might then be used to estimate the clustering amplitude of GRB absorbers.

\section{Statistics of absorption lines and beam size}
\label{beamsize}
Frank et al. (2006, herafter F06) proposed a geometrical explanation for
the different incidence rate of strong \mg systems in QSO and GRB lines
of sight.
They argued that the difference in the statistics
can be readily explained if the size of the QSO beam is
approximately 2 times larger than the GRB beam, provided the
size of the \mg absorbers is comparable to the GRB beam and of the order
of $\mincir 10^{16}$ cm$^{2}$. 
This is in contrast with a number of theoretical and observational
estimates suggesting that the typical QSO beam is a few times
smaller than the GRB beam (see Section 4 in F06 and references therein). 
We show here that the solution worked out by F06 did not include an
additional effect which changes the outcome of their model.

In what follows $r$ is the comoving radial distance and $D_{\rm a}$ is
the angular diameter distance. For simplicity, let us consider a
population of gas clouds with comoving number density $n(z)$, proper
cross-section $\sigma$ and equivalent width distribution $f(W_0)$
(normalized to unity).
The clouds can be detected as absorption lines in the
spectrum of background continuum sources with angular beamsize $\Omega_{\rm
b}$. 
If the beamsize is
much smaller than the solid angle subtended by the gas clouds,
i.e. $\Omega_{\rm b}\ll \sigma/D_{\rm a}^2(z)$, the number density of
absorption lines per unit redshift and equivalent width is 
\be
\frac{d^2N}{dW_0\, dz} =\sigma\,n
%(z)
\,(1+z)^2\,\frac{dr}{dz}\,f(W_0)\;.  
\ee 
On the other hand,
if the clouds are much smaller than the beam of the background source,
i.e. $\Omega_{\rm b}\gg \sigma/D_{\rm a}^2(z)$, two new effects need to be
accounted for. First,
the observed equivalent width is proportional to
the covering factor of the cloud (i.e. to the fraction of the beam area
covered by a single absorber) $W_{\rm obs}=
\sigma/[\Omega_{\rm b} D_a^2(z)]\, W_0=
[\sigma/\sigma_{\rm b}(z)]\, W_0$.
Second, the mean number of absorbers scales proportionally to the
area of the beam
%\be
%\frac{dN}{dz} 
%(z)
%=\sigma_{\rm b}\,n(z)\,(1+z)^2\,\frac{dr}{dz}\;
%\ee
(note that this effect was apparently neglected by F06) so that:
\be
\frac{d^2N}{dW_{\rm obs}\, dz}
% (W_{\rm obs},z)= 
=\sigma_{\rm b}\,n
%(z)
\,(1+z)^2\,\frac{dr}{dz}\,\frac{\sigma_{\rm b}
%(z)
}{\sigma}
\,f\left(\frac{\sigma_{\rm b}}{\sigma}\,W_{\rm obs}\right)\;.
\label{dis1}
\ee
The last possibility is obtained when $\sigma_{\rm b} \simeq \sigma$.
In this case, the number of absorption lines is 
\be \frac{dN}{dz}\simeq
(\sqrt{\sigma}+\sqrt{\sigma_{\rm b}})^2\,n
%(z)
\,(1+z)^2\,\frac{dr}{dz}\;, \ee and 
Monte Carlo simulations 
must be used to derive the corresponding distribution
of $W_{\rm obs}$ that depends
on the relative positions and shape of beam and absorbers.

Let us now consider two different classes of sources (for instance GRBs
and QSOs), and denote their relative beamsize by $x=\Omega_{\rm
G}/\Omega_{\rm Q}$.  
At every redshift,
from equation (\ref{dis1}) we derive \be
\frac{d^2N_G}{dW_{\rm G}\, dz} (W_{\rm G},z)=x^2
\,\frac{d^2N_Q}{dW_{\rm Q}\, dz}\left(W_{\rm Q}=xW_{\rm G},z\right)\;. \ee 
Therefore, to
find four times more absorbers with $W>1$ \AA \ along GRBs than along
QSOs, the relative beam size must satisfy the equation $4 \int_{1{\rm
\AA}}^{\infty} \frac{d^2N_Q}{dW_{\rm Q}\, dz}\,dW_{\rm Q}
=\int_{1{\mathrm\AA}}^{\infty} \frac{d^2N_G}{dW_{\rm G}\, dz}\,dW_{G}=
x\int_{x{\mathrm \AA}}^{\infty} \frac{d^2N_Q}{dW_{\rm Q}\, dz}\,dW_{\rm
Q}$.  Adopting the fit for the equivalent width distribution derived 
either by Nestor et al. (2005) or by Prochter et al. (2006a) 
it is easy to show that this equation
does not admit solutions. 
\begin{figure}
\plotone{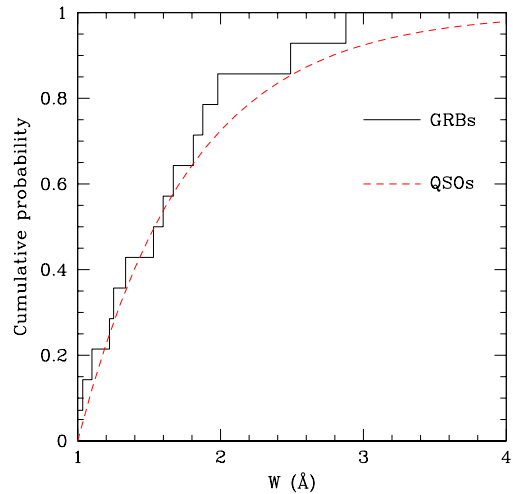}
\caption{\label{ks}
The cumulative probability distribution of the \mg equivalent widths along GRBs listed in P06 (solid histogram)
is compared with the fit for QSO absorbers given by Prochter et al. (2006a). The two distributions are compatible
at the 95.7 \% confidence level in the Kolmogorov-Smirnov sense.}
\end{figure}

This applies to small absorbers and only
approximates the case in which 
absorbers and beams have similar sizes. 
We performed a number of Monte Carlo simulations varying the beam
and absorber sizes. For GRB afterglows we both considered disk and 
ring geometries (with varying thickness)
while we only considered disk-like QSO beams. In no case 
could we reproduce the observational results. 
If the F06 solution holds, the distribution of \mg equivalent widths 
along GRBs should be flatter than in QSOs (this is also evident in our
Monte Carlo simulations). 
However, we found that
the equivalent widths listed in P06 are perfectly consistent 
with the distribution derived from SDSS QSOs (Figure \ref{ks}).
The Kolmogorov-Smirnov test suggests that the observed equivalent widths 
for GRB absorbers are compatible (at the 95.7 \% confidence level) with
being a random sampling of the probability distribution derived from SDSS QSOs. %

There are some other observational results that the solution proposed
by F06 cannot explain: {\it i)} along QSOs there are
no apparently unsaturated \mg absorption lines
with a doublet ratio (1:1), as would be expected if small 
saturated systems were being diluted by a larger QSO beam;
{\it ii)} an apparently unsaturated
\mg absorption system with a doublet ratio (1:1) has been detected
along GRB030226 (Shin et al. 2006) in the spectrum of GRB030226 thus
suggesting partial covering of the GRB beam; {\it iii)} if the strong \mg systems resemble
intercloud medium of the MW their sizes are likely to be in the range
1-1000 pc (e.g. Kobayashi et al. 2002; Rauch et al. 2002; Churchill et
al. 2003; Ding et al. 2003) which are difficult to reconcile with the
sizes proposed by F06.

\section{Statistics of absorption lines and lensing}
\label{lensing}
Only a few spectra of optical GRB afterglows have been taken.
Is it possible that the sample is heavily biased towards lines
of sight intersecting
a large number of absorbers?
Gravitational lensing due to intervening material
could in principle boost the afterglow luminosity
by amplifying its beam size. This would make optical spectra easier to take.

Combining the data by P06 with those by Nardini et al. (2006), we find 
that afterglows with more than one \mg absorbers
are, on average, 
a factor of 1.7 brighter\footnote{In terms of their optical
luminosity measured 12 hours (in the GRB rest frame) after the trigger.}
than the others. According to a t-Student test,  
there is a 10\% chance that this difference is due to random fluctuations.
If this result is strengthened by increased statistics
(currently only four afterglows showing more than one strong \mg absorber are
known),
it is then likely that some form of lensing caused by mass concentrations 
associated with the absorbers themselves 
is the cause of their increased detection probability.

The galaxies hosting the absorbers represent obvious lens candidates.
However, since \mg absorbers are seen at large distances from galaxy cores,
it is unlikely that they can produce large magnifications.
Our Monte Carlo simulations presented in \S\ref{dust} show
that lines of sight with more than 2 \mg absorbers have a mean magnification
of 1.04 and only 2\% of them are magnified by more than a factor of 1.2
(see also M\'enard 2005).
Strong lensing is a very rare phenomenon. Only a few GRBs 
in every thousand detections are expected to be strongly lensed by galaxy-sized
halos (Porciani \& Madau 2001). 
Millilensing by $10^{7} M_{\odot}$ haloes has an optical depth which is
about 1000 times smaller than the strong lensing one (Porciani \& Madau
2000).

Afterglow spectra are generally taken within the first few 
hours after the gamma-ray triggering event when most of the optical emission
is expected to come from a narrow ring of radius $r_{\rm s}\simeq 
4\times 10^{15}\,(t/1\, {\rm hr})^{5/8} (1+z_{\rm GRB})^{-5/8}$ cm 
which is expanding at superluminal 
speed on the sky (Waxman 1997).
%superluminal
%expansion at a speed $\sim \gamma c$, with $\gamma$ the Lorentz factor of the 
%GRB shock (Waxman 1997).
The beam size is thus comparable with the Einstein radius
of compact solar mass objects at cosmological distances $r_{\rm E}\simeq
5\times 10^{16} (M/M_\odot)^{1/2}$ and GRB afterglows can be efficiently
microlensed by intervening stars and massive compact objects (MACHOs).
When the source can be regarded as pointlike with respect to the lens, 
microlensing produces a constant amplification depending on the impact
parameter of the lens $b$ (Loeb \& Perna 1998). 
On timescales of days (observer frame),
the magnification then increases and reaches 
a maximum value when the source size crosses the lens ($r_{\rm s}=b$). 
This rather sharp brightening is a characteristic signature of microlensing
events. For large impact parameters, when the mean magnification is $<2$, 
and for broader source rings
this feature is less pronounced (see Figure 1 in Loeb \& Perna 1998)
and could then be not easily detected observationally.
For sources at $z\sim 2$,
the microlensing optical depth is $\tau\simeq 0.65\, \Omega_{\mu{\rm l}}$
where $\Omega_{\mu{\rm l}}$ denotes the present-day density of microlenses
in units of the critical density of the Universe (Baltz \& Hui 2005). 
Even in the most optimistic assumption that 20\% of the dark matter is in
MACHOs, only a few percent of the GRB afterglows 
should be affected by microlensing.
In summary, unless the intrinsic luminosity function of GRBs (and
of their afterglows) is extremely steep and magnification bias plays
an important role (see below), 
a lensing explanation of the increased \mg absorption along GRB 
lines of sight 
should be regarded as very unlikely in the standard cosmological scenario.

The lensing solution works better in the presence of a cosmological
population of small dark matter clumps.
Mini clusters of
dark matter (axion-like or Higgs-like) particles naturally form when a scalar field undergoes
a second order phase transition below the QCD scale (Hogan \& Rees 1988). 
Their existence does not violate any observational constrain provided that their
mass is smaller than $10^4$ M$_\odot$  and 
they would behave as efficient lenses if their physical size is smaller than their Einstein ring,
which corresponds to a lower mass limit of a few $\times 10^{-6}$ M$_\odot$ (Zurek, Hogan \& Quinn 2006).
Assuming that all the dark matter is in mini-clusters, we expect 
%and extrapolating the results of Baltz \& Hui (2005), we find 
that nearly $30\%$ of the {\it Swift} afterglows could be strongly microlensed.
 %Note that
%the variability time scale for these lensing events is of the order of 1 yr 
%and
%thus they would not be tagged as special events from observations of the optical afterglow.
%
To explain the observed abundance of \mg absorbers, however, QSOs need
not to suffer from these microlensing events. In other words, QSO
beams have to be larger than the Einstein ring of the mini clusters
(and, thus, of the GRB beams). Current estimates of the QSO size
($<10^{15-16}$ cm) then favour mini cluster masses which are smaller
than 1 M$_\odot$.  Patchy obscuration of the GRB optical beam by dust
in the circumburst environment could make GRB beams smaller than
expected.
Albeit very speculative,
this scenario would certainly favour the detection of afterglows spectra with 
an increased number of absorbers 
even though it is difficult to estimate the net effect on the observed
$dN/dz$. 
Note that this result is not in contradiction with Section 4. 
If the absorber size is much larger than both the beam sizes then the
equivalent width distribution function is
expected to be the same for QSOs and GRBs.

%has been
%investigated in great detail by Baltz \& Hui (2005). 
%Under the most optimistic assumptions ($20\%$ of dark matter
%in MACHOs), they found that only 1/27
%of the GRB spectra observed by {\it Swift} is likely to be
%microlensed. 

An appealing possibility
is that GRB afterglows are strongly affected by 
magnification bias.
Light sources which are particularly bright in more than one waveband
are especially likely to be lensed
(Borgeest, von Linde \& Refsdal 1991; Wyithe, Winn \& Rusin 2003).
This ``multiband magnification bias'' could be very important 
for GRBs whose optical and gamma-ray luminosities are found to be
statistically independent (Nardini et al. 2006). 
It is easy to show that, independently of the lensing optical depth (and
thus of the kind of lens),
the fraction of lensed objects at a given redshift
approaches unity when the mean faint-end slope of the luminosity
functions in the two bands approaches -2 (Wyithe et al. 2003).
\footnote{The detection of an X-ray signal is often required to accurately
locate GRB afterglows for optical follow up. Three-band magnification bias 
is even more efficient than the case discussed in the main text. For 
independent luminosities in the different bands, the critical mean faint-end
slope is -5/3 (close to the observed value for the prompt gamma-ray emission).
Note, however, that gamma and X luminosities of GRB afterglows seem to 
correlate rather tighly (e.g. Nardini et al. 2006) and this would again imply
a critical slope closer to -2.}
For the faint-end of the gamma-ray luminosity function,
the universal structured-jet model predicts a slope of $\gamma=-2$
(Rossi et al. 2002; Zhang \& M\'esz\'aros 2002) while
observational estimates based on number counts give
$\gamma=-1.57\pm 0.03$ (Firmani et al. 2005) 
and $\gamma=-1.7\pm 0.1$ (Schaefer, Deng \& Band 2001).
The luminosity function of optical afterglows is not known. 
Nardini et al. (2006) find that most of the observed 
afterglows have a similar luminosity but they do not discuss
selection effects.
The fact that many afterglows are not detected in the optical band 
might suggest that the slope of the luminosity function
is rather steep indeed.
Magnification bias could then be responsible of the different counts
of \mg absorbers between GRBs and QSOs.
Note that optically selected QSOs have a faint-end slope of $\gamma\simeq
-1.6$ (Boyle et al. 2000; Croom et al. 2004) while (single-band)
magnification bias becomes important for $\gamma\simeq -3$, thereby under this scenario we do not expect
QSOs to be affected by microlensing events.

\section{Association of \mg systems to GRBs}
\label{association}
We now consider the possibility that the strong \mg systems are
associated with the GRB event. This solution is particularly
attractive since it gives an ``additive'' correction instead of a
``multiplicative'' one.  In other words, at variance with the
solutions explored so far that multiply the incidence rate of \mg
lines by a constant factor, it is sufficient that a few of the 14 P06
absorbers are associated to the GRB events to fully solve the
discrepancy in $dN/dz$. Both the hint for a partial covering of \mg
systems along some GRB lines of sight and the failed optical detection of the
galaxies responsible for \mg absorptions (Ellison et al. 2006) support
this hypothesis.
The major limitation, however, is the need for cold, metal enriched gas
moving at semi-relativistic speeds.  Even assuming that the
five P06 systems with the largest inferred velocities are produced by
intervening galaxies (as expected from \S \ref{dust}), while the rest
are associated with GRB events, we still find that the mean peculiar
velocity of the intrinsic systems is $6 \times 10^4$ km s$^{-1}$ (with a
dispersion of $3\times 10^4$ km s$^{-1}$).

In the fireball model, 
afterglow emission is produced when the fastest ejecta from the central
engine sweep up and shock the circumstellar medium. This happens at typical
distances of $10^{16-17}$ cm from the central engine.
Gas responsible of \mg absorption must thus lie further away than this.
In order to produce absorption with column densities of $10^{15}$ cm$^{-2}$ which
fully cover a QSO beam one needs as little as $10^{-7}$ M$_\odot$ of material
with solar metallicity.
Wolf-Rayet winds, binary interactions and a SNa explosion could have
easily transported such an amount of metal enriched gas at these distances.
The GRB shock has enough energy to accelerate this gas to semi-relativistic
velocities only within $\sim 10^{18}$ cm from the central engine.
However, most likely, the swept up material will also be heated to temperatures
well above the ionization threshold of \mg. 
\footnote{Unless the relativistic shock does not penetrate the dense, 
metal-rich cloudlet, which is also possible (M. Vietri personal communication)
and would make \mg absorption easier.
}
Therefore no \mg absorption will be possible
until the gas temperature cools down below $\sim {\rm a\ few}\ \times 10^4$ K.
Even though the action of
Rayleigh-Taylor instabilities will help reshaping the structure of the clouds and produce a multi-phase
medium, it seems unlikely that this sequence of phenomena could produce
the observed absorption.
It is also difficult to conceive how \mg systems could appear 
in the intense photoionization field of the GRB afterglow. 
%where
%$10^{59-60}$ photons per steradian per day can potentially
%produce a second ionization of Mg atoms.
%
%
The characteristic isotropic equivalent luminosity 
of the optical afterglow in the Cousins
R-band after 12 hours (in the rest frame) 
is $4.47\times 10^{30}$ erg s$^{-1}$ Hz$^{-1}$ (Nardini et al. 2006).
Assuming a power-law spectral energy distribution with index $\beta$
and a time decay with index $\alpha$, the number of ionizing photons 
(with energy above the second photoionization threshold for Mg, 
$E>E_{\rm thr}=15.04$ eV) emitted per unit solid angle 
from $t_{\rm min}$ to $t=t_{\rm max}$ (both in seconds) is
\begin{eqnarray}
\frac{dN_{\rm phot}}{d\Omega}&=&\frac{5.37\times 10^{55}\,t_{\rm max}}{\beta\, (1-\alpha)}\,
 \left(\frac{1.82\, {\rm eV}}{E_{\rm thr}}\right)^\beta \nonumber\\
&\times &\left(\frac{43200}{t_{\rm max}}\right)^\alpha\, 
\left(1-\left(\frac{t_{\rm min}}{t_{\rm max}}\right)^{1-\alpha}\right)\;.
\end{eqnarray}
For typical values $\alpha=\beta=1.3$, $t_{\min}=$ 3 days,
$t_{\rm max}=$ 1 month, this gives 
$10^{58}$ photons per steradian that can potentially
produce a second ionization of Mg atoms.
Most of these photons will be actually absorbed by atoms of H,C,N and O
which (assuming solar abundance ratios) 
are more abundant than Mg and have lower photo-ionization thresholds.
An optically thin shell, $n<20$ nuclei per cm$^{3}$, exposed to such
a photo-ionizing flux gets fully ionized on extremely short
timescales. Detailed calculations for higher density clouds require an accurate
treatment of radiative transfer and are beyond the scope of this paper.
The presence of structured jets (with fast and slow components), 
already invoked to explain
\CIV, \SiIV and \HI absorption associated with GRB 021004
(Starling et al. 2005), is probably needed.

In the less favoured supranova model (Vietri \& Stella 1998), a supernova event preceeds
the GRB with a time lag ranging from weeks to a year.
Radio observations suggest that the fastest ejecta of ``normal'' supernovae 
Ibc (i.e. fast electrons)
have typical velocities of $\sim 0.3 c$ (Berger et al. 2003)
which are in agreement with the velocities of the absorbers in P06.
If the ejecta travel at a constant velocity of $\sim 10^5$ km s$^{-1}$,
they will need $\sim10-100$ days to cover a distance of $10^{16-17}$ cm.
Longer delays are needed to transport the metals produced by the supernova
at these distances.
The observed narrowness of \mg lines ($\Delta v \sim$ a few
hundred km s$^{-1}$) poses some strong constraints on the physical size of
the absorbers.  
To first approximation, in an explosive event, $\Delta
v/v=\Delta R/R$ (with $v$ velocity of the ejecta, $R$ their distance
from the supernova) which implies a linear size for the absorbers of $\sim
10^{12-13}$ cm. Thereby, these systems could either
be associated with narrow and dense shell-like structures
or to small condensations.
However, it is difficult to imagine how the gas could keep cold
and to reconcile the required delays between the supernova and the GRB phases 
with observations (e.g. M\'esz\'aros 2006).

An alternative mechanism for producing the observed \mg system requires that
the material emitting the afterglow is lumpy.
M\'esz\'aros \& Rees (1998) investigated spectral features arising from ultrarelativistic
ions in the pre-afterglow  fireball outflow. In this model, metal-rich
inhomogeneities (blobs or filaments) are 
typically very small ($10^5$ cm) and dense ($\sim10^{18}$ cm$^{-3}$), 
although their
surface covering factor can be quite high. If these structures survive
up to the afterglow emission they could be responsible of the
observed \mg absorption since they are entrained in the flow that produces
the burst. Some kind of magnetic mechanism should be, however, invoked
to keep these blobs collimated in the afterglow phase. 

A promising scenario is to assume that the observed \mg systems are associated
with supernova remnants (SNRs) in the vicinity of the GRB.
MgII absorption lines from local SNRs and superbubbles look similar to those detected along high-$z$ QSOs 
and appear to be made of several components with small velocity
dispersion (Danks 2000, Bond et al. 2001, Welsh et al. 2001).
Imagine that a GRB line of sight crosses the shell of a young SNR that is expanding at a velocity of $v\sim 3\times 10^4$ km s$^{-1}$.
Also assume that the shell is spherical and that it produces two absorption features
(we postulate the presence of cold pockets of gas in the shell)
in the GRB spectrum: one blueshifted and one redshifted by the amount $(1+z_{\rm GRB})(1\pm v/c)$.
Since GRBs are normally associated with the highest-redshift absorbers in their spectra,
one would erroneously assign a redshift $(1+z_{\rm GRB})(1+v/c)$ to the GRB and
say that the second absorber moves with a relative velocity of $2v$ with respect to it.
How likely is it that a GRB lines of sight intersects a SNR?  A typical
star forming region like 30 Doradus has a characteristic size of $\sim
100$ pc, while a young SNR has a radius of $\sim 10$ pc.  The optical
depth for crossing a shell along a random line of sight towards the GRB
is thus of order unity if the number density of SNRs is $\sim 3\times 10^{-5}$
pc$^{-3}$, which corresponds to nearly 15 objects within a single star
forming region.  X-ray studies of 30 Doradus have detected several SNR
candidates and five superbubbles (Townsley et al. 2006 and references
therein) thus showing that the probability for a line of sight to cross a SNR
is not negligible.
Note that not in all GRB spectra the expanding shell of a SNR will produce 
two absorption lines. Sometimes features of the host galaxy will
be present together with one line from the SNR. In this case the
velocity difference between the lines will be smaller.

\section{Summary}
\label{summary}
We addressed possible explanations for the different
incidence rate of strong \mg systems along GRBs which is about $\sim 4$
times higher than along QSOs. 
%
%
%Even though we could not single out a physical effect that fully accounts for
%the observed counts, 
There are a number of phenomena that potentially 
contribute: dust-obscuration bias, different beams sizes between GRBs
and QSOs, gravitational lensing, association of GRB absorbers with
the circumburst environment. Our results can be summarized as follows.
\begin{itemize}
\item
Considering a simplified model for \mg absorbtion based on 
a set of realistic numerical simulations of galaxy formation
and able to reproduce all the observational constraints,
we showed 
that the incidence rate of \mg systems in QSO spectra can be
underestimated by a factor 1.3-2 due to dust obscuration bias.
This is not enough to fully explain the discrepancy with the number
of absorbers along GRBs.
\item
We critically discussed the solution proposed by F06 where
QSOs are assumed to be a factor of two larger than GRBs. Accounting for
the dependence of 
both the equivalent width and the number density of absorbers with the beam
size, we showed that it is not possible to fully explain the difference 
in the observed counts by assuming that the two classes
of background sources have different characteristic sizes.
We also found that the equivalent-width distribution of \mg systems detected
along GRB lines of sight is compatible with being a random sampling of
the QSO one at the 95.7 per cent confidence level.
This suggests that the absorbers are larger than the sources and provides
evidence against the solution proposed by F06.
\item
We showed that GRB afterglows with more than one absorber
are brighter than the others by a factor of 1.7. 
If confirmed by increased
statistics, this finding would suggest a lensing origin of the 
\mg discrepancy.
However, in the standard cosmological scenario, 
lensing optical depths are small and difficult to reconcile with
the observed counts of \mg absorbers.
This may hint towards the existence of an exotic population of lenses.
For instance,
microlensing due to mini dark matter clumps could explain the difference
but this also requires QSO beams to be larger than GRB afterglows which
is in disagreement with current observational estimates.
\item
Due to their multiband selection and their emission properties, 
GRB afterglows are particularly 
sensitive to magnification bias. 
Independently of the nature of the underlying lenses,
a large fraction of the observed GRBs would be lensed 
if the
average faint-end slope of their gamma and optical luminosity functions
approaches -5/3 -- -2. 
According to current models for prompt and 
afterglow emission this is not unlikely. Present observations suggest a
slope of $\gamma\sim -1.6$ for the prompt gamma-ray emission+
while the luminosity functions of
optical afterglows is totally unconstrained.
Magnification bias then provides a viable solution to the observed
discrepancy.
\item
The production of
\mg absorbers in the circumburst environment 
(either by cold and dense cloudlets of metal-rich material 
accelerated to semi-relativistic speeds
or, most likely, 
by SNRs unrelated to the GRB event itself but lying in the
same star forming region) could further reduce
the statistical significance of the observed discrepancy to a level
compatible with statistical fluctuations due to small number statistics.
\end{itemize}

We conclude that, 
with the possible exception of magnification bias, it is unlikely
that one of these effects on its own can fully account for
the observed counts. However, the combined action of some of them can
substantially reduce the statistical significance of the discrepancy
between the incidence of strong \mg systems in QSO and GRB spectra.

\acknowledgments We thank B. Carswell, M. Haehnelt, F. Miniati,
M. Pettini, E. Pian, M. Rees and M. Vietri 
for useful discussions. P. Madau is warmly thanked for comments on an
early version of this paper. MV thanks the
hospitality of ETH where this project was started.

{}

\end{document}